\documentclass[10pt, sigconf]{acmart}

\usepackage{booktabs} 
\usepackage{amsmath,amssymb,amsfonts}
\usepackage{algorithmic}
\usepackage{graphicx}
\usepackage{textcomp}
\usepackage{xcolor}
\usepackage{amsmath}
\usepackage{subcaption}
\usepackage{diagbox}

\setcopyright{rightsretained}

\acmDOI{}

\acmISBN{}

\acmConference[]{}{}{}
\copyrightyear{2018}

\begin{document}
\title{A Learning Approach to Wi-Fi Access}

\author{Thomas Sandholm}
\affiliation{
  \institution{CableLabs Core Innovation}
  \city{Sunnyvale}
  \state{California}
}
\email{t.sandholm@cablelabs.com}

\author{Bernardo A. Huberman}
\affiliation{
  \institution{CableLabs Core Innovation}
  \city{Sunnyvale}
  \state{California}
}
\email{b.huberman@cablelabs.com}

\begin{abstract}
We show experimentally that workload-based
AP-STA associations can improve system
throughput significantly. We present a predictive
model that guides optimal resource
allocations in dense Wi-Fi networks and achieves 72-77\% of the
optimal throughput with varying
training data set sizes using a 3-day trace of real cable modem traffic.
\end{abstract}

%
%



\maketitle

\section{Introduction}
The rapid rise in deployments of connected IoT and AI devices, as well as the
continued growth in adoption of mobile devices, has lead to a new surge in
Wi-Fi usage. At the same time, more Wi-Fi access points (APs) are appearing to off-load
cellular usage, share cellular connections across devices, and to extend signals
in mesh networks, or simply to serve more businesses and local stores.
Enterprises and large conferences have long battled 
the problem of serving many stations from a pool of access points efficiently,
and it is well known that simply directing stations (STAs) to the closest access points
can be suboptimal~\cite{murty2008}. The latest 802.11ac and 802.11ax standards
help serve this demand by bonding channels and allowing wider bands to improve
throughput. The use of wider bands, however, introduces more side lobe interference and
leads to fewer orthogonal bands.




A study of a campus WLAN (\cite{balazinska2003characterizing}) concluded that
user transfer rates follow a power law, and as a result, load tends to be unevenly 
distributed across access points. Furthermore, they discovered that which users are active
is more significant to the load incurred than how many users are connected. This finding motivates
us to look at the problem of load balancing from a STA workload perspective. Many studies
have shown that traditional STA to AP association driven by the wireless clients 
can be suboptimal and that AP-driven central control is beneficial~\cite{zeljkovic2017exploiting,sheshadri2016,lin2017two}.

The problem we are addressing in this paper is the following. When STAs 
have a choice of which AP to connect to and each AP
has a fixed independent capacity constraint on the bandwidth offered, how do we allocate
the aggregate bandwidth most efficiently?

Distance to the AP clearly plays a role, but we argue that the characteristics of the workload
play an even bigger role in dense deployments. With a very large number of independent random workloads one can
expect statistical multiplexing to even out the peaks and valleys in usage and result in
little to no unused resources. However, given the limited frequency band widths, 
having too many stations on the same frequency is inefficient. 

Stations in general outnumber access points, which use a smaller contention window
to adjust the backoff algorithm to serve more stations efficiently. This typically
results in two download streams (AP to stations) receiving a higher throughput
than mixing an upload and a download stream on the same AP. Similar effects can
be seen when mixing low latency applications with high-throughput ones.
As a result, dual radio approaches have been suggested~\cite{edgewater2018}, to separate out 
different application streams on different channels.

Our work focuses on performing this separation automatically based on observed
workload rates.

Two use cases motivating heterogeneous bandwidth scheduling are community Wi-Fi
sharing~\cite{powell2008} and channel bonding~\cite{sun2014}. These use cases correspond to the backhaul capacity
and the airtime bandwidth being scarce (and non-uniform) respectively. 
Community Wi-Fi is becoming increasingly popular to increase coverage and provide
an open alternative to cellular networks, whereas channel bonding is the key strategy
(together with multi-antenna solutions like MIMO) that the
most recent 802.11 specifications~\cite{gast2013,aguilera2018} apply to address increased throughput demand. 

The key contribution of this paper is an experimental evaluation of workload-aware
STA-AP associations in a dense network setting for both backhaul and airtime
constrained systems. To the best of our knowledge, this is
the first experimental evaluation of such associations, using predictive, 
demand-exploring algorithms.

\section{Related Work}\label{sec:relatedwork}
A large body of research has addressed the problem of
load imbalance due to suboptimal STA to AP associations.
The work can be categorized into: {\it beacon-based decentralized
approaches}, modifying the protocol of association between
the AP and STA; {\it centralized load balancers}, making association decisions
from a controller with
a global view of the network;
{\it virtual radios}, separating the AP into a dumb radio and
a centralized virtual AP that makes association decisions; and finally {\it latency
aware approaches}, balancing the resources between throughput and
latency sensitive applications.

{\bf Decentralized Beacons.} Measurement beacons are used in ~\cite{vasudevan2005} to 
predict effective throughput before
connecting to an AP. 
\cite{gong2008dynamic} extend beacons with load that STAs can react to. The
load information transmitted is aware of the adaptive data rates selected by the AP. 
The key feature of these approaches is that they are distributed so as to adapt to
changes in load gradually. A similar approach using admission control
and custom AP load functions is presented in~\cite{kim2012distributed}.
A cell breathing algorithm, adjusting the power of beacon packets
of APs to effectively shrink or grow the cell covered by an AP
dynamically based on load was evaluated in ~\cite{bahl2006}.
In \cite{balachandran2002hot} channel switching to nearby APs during overload
is used as a remedy for imbalance whereby APs force STAs to switch.
\cite{papanikos2001study} implement an AP load balancing strategy that combines
controlled channel selection, number of users per AP, and link quality again by extending
the information sent in probe responses from APs to STAs. Our approach
does not require any modification to the protocols neither on the AP nor the STA side,
and can make more informed association decisions thanks to a central view of the network.

{\bf Central Controllers.} Optimal STA to AP association based on
expected wait times on the MAC level were suggested in ~\cite{broustis2010}.
AP load balancing based on free airtime measurements and
interference from nearby clients and APs were also considered in the
DenseAP system presented in~\cite{murty2008}.
In~\cite{bejerano2004fairness} the authors address the problem of
fairness across users accessing a wireless LAN by association control, i.e.
which STA is mapped to which AP. They show in simulations that their proposed
association algorithm achieves close to optimal load balancing and max-min fairness.
Centralized Radio Network Controllers were also explored in
~\cite{vasan2005}. Architecturally, this category of approaches is closest to our work.
We extend the existing work by incorporating information about and predicting the workload
generated by individual STAs over time. Given the
architectural similarity we envision our method to be deployed either as a compliment
or a replacement in existing WLAN controllers.  

{\bf Virtual Radios.} In ~\cite{sheshadri2016} the RF transmission (radio) part
of the AP is separated out and the baseband processing is
served in a clustered pool of compute resources connected
with high-speed fiber. The architecture allows real-time
mapping of channels and radios to APs based on traffic load. Similar SDN-based
approaches were also exploited in~\cite{zeljkovic2017exploiting} and~\cite{lin2017two}.
These approaches are very similar to the centralized approaches with the
main difference being that the central controller has even more information and
control over the APs and that the handovers can be executed with less
overhead. Our approach also relies on smooth handovers and could be deployed
as a compliment to virtual radio association algorithms. The additional information
in the centralized AP makes it easy to run our STA workload aware algorithm.

{\bf Latency Aware.} \cite{nishat2016} adapt
channel width and move stations between channels dynamically
based on observed frame rates to
allow both latency and throughput sensitive applications
to be served at a high QoS concurrently. Similarly~\cite{das2017}
optimises low latency streams by using a ping protocol to 
determine whether delays can be explained
by cross-traffic or the station's own transmission bitrate to avoid
backing off too conservatively to avoid congestion. In~\cite{chakraborty2016}
{\it experiental capacity regions} are proposed that represent mixes of STAs that
can be admitted into a wireless network
based on QoE capacity requirements using a machine learning model.
Our work currently focuses on bandwidth scheduling but
more general applications to also pack STAs on APs based on latency
requirements fits well with our general architecture and is future work.

\section{Model}

The general problem we address is how to best associate 
STAs (wireless clients) with APs (wireless access points). In
this work, the best allocation is the one that yields the
highest overall system throughput. Each AP has a fixed 
transmit and receive capacity, translating to maximum 
achievable upload and download rates from the STA point of view.

The key variable in this setting is the upload and
download rates generated over time by the STAs.

We define an observation window to be one or more
time slots in the past for which we know the 
upload or download rates across all existing STAs.

Similarly we define an allocation window to be one
or more time slots in the future for which an
allocation is executed.

An {\it allocation} here is an instance of a AP-STA association
connecting each STA to an AP. We call all feasible 
allocations (associations where STAs are within reach of APs)
the feasible allocation set.

Now, the problem to be solved is the following:
Given a set of observations in a historical time window, provide
a prediction of the best allocation for the future
time window out of the feasible allocation set.

Given that the allocations are enforced in the AP, or a controller, we assume that the
feasible allocations are known. 

Our proposed model predicts future demand using correlations between
observed workload rates and optimal allocations.

We use a set of linear regression models (LR) to predict the throughput
of a feasible allocation. Each model uses observed download and upload 
rates in a particular state of the system. The state is simply defined
as the current allocation being enforced. We then define the score, $S$, 
of a feasible allocation $A'$ given an observed state, $A$, as follows:
\begin{equation}
	S_{A,A'} = \sum_{s=1}^N {w^d_s r^d_s + w^u_s r^u_s} 
\end{equation}
where $w$ represents the regression model coefficients to be learned for each station, $s$,
of the $N$ stations, $r^d$ the download rates and $r^u$ the upload rates observed during
allocation $A$. 
The score here is the same as the predicted throughput, and hence
we pick the allocation $A'$ with the highest score in the current state to enforce,
in order to optimize overall system throughput.
This means that we train a different model for each allocation used in the observation window, and 
each model for which we want to predict the throughput. The total number of linear models to train is
hence $|A|*|A'|$. 
Note, we do not need to observe the same workload demand with all allocation models 
in each time period in order to train the models. For example, with two allocations $\{a,b\}$ 
we can observe rates with allocation $a$ in time $t$, then get ground truth throughput in $t+1$, as well as observe the new rates. Then we change to
allocation $b$ in time periods $t+2$ and $t+3$ to repeat the procedure, and then finally
switch back to allocation $A_a$. In this way we have trained all of our four models $\{S_{a,a}\}$,
$\{S_{a,b}\}$, $\{S_{b,b}\}$ and $\{S_{b,a}\}$ in five time steps. 

\section{Data}
Given that the model is predictive and should be capable of learning some hidden
behavior in the upload and download rates, we collect a real-world trace from
a residential deployment with cable modems connected to a cable headend (CMTS)
over a HFC network. The data comprise upload and download volumes on a per-second basis
for each cable modem that we aggregate into upload and download rates on a
minute-by-minute basis. Here, we consider each cable modem trace a proxy for
a workload even though it may be a combination of many wireless and wired
stations in the home.

Three days of rates were captured
from July 5th through July 7th 2017 from 8~\footnote{we have data from 21 modems but only use 
the top 8 in terms of traffic volume for the primary experiments} cable modems. 

\begin{figure}[htbp]
        \centerline{\includegraphics[scale=0.5]{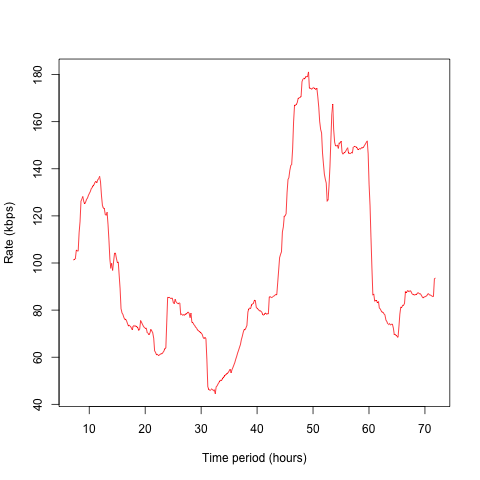}}
\caption{Aggregate Average Upload Rates over 10min Periods.}
\label{upload}
\end{figure}

\begin{figure}[htbp]
        \centerline{\includegraphics[scale=0.5]{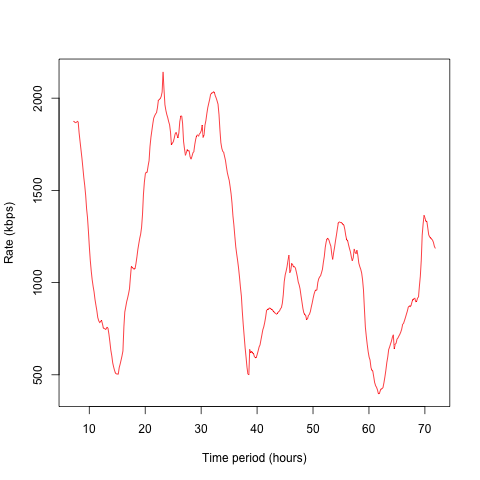}}
\caption{Aggregate Average Download Rates over 10min Periods.}
\label{download}
\end{figure}

The aggregate average upload and download rates with smoothing
over a 10min period can be seen in Figure~\ref{upload} and
Figure~\ref{download} respectively.

\begin{figure*}
	\centering
	\begin{subfigure}[b]{0.24\textwidth}
		\includegraphics[width=\textwidth]{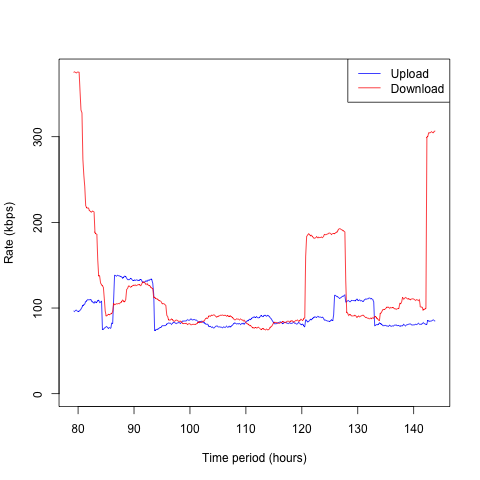}
	\end{subfigure}
	\begin{subfigure}[b]{0.24\textwidth}
		\includegraphics[width=\textwidth]{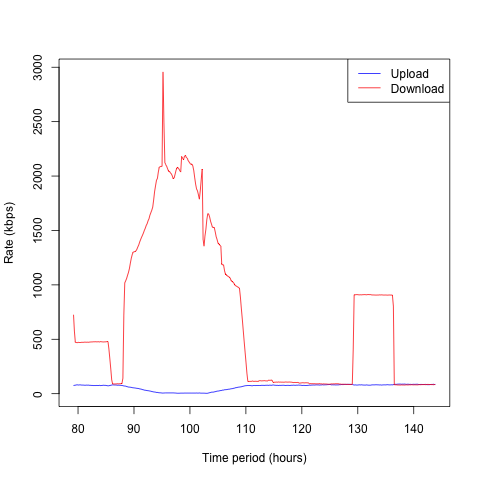}
	\end{subfigure}
	\begin{subfigure}[b]{0.24\textwidth}
		\includegraphics[width=\textwidth]{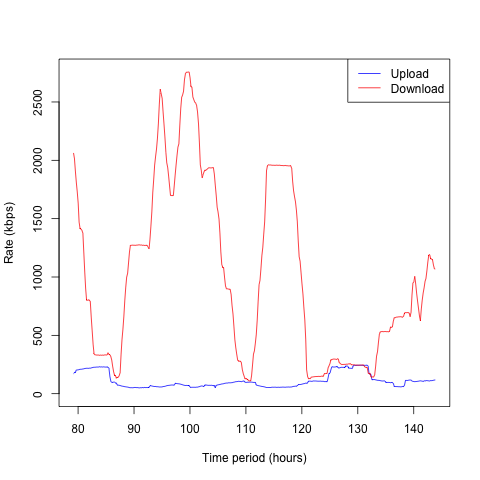}
	\end{subfigure}
	\begin{subfigure}[b]{0.24\textwidth}
		\includegraphics[width=\textwidth]{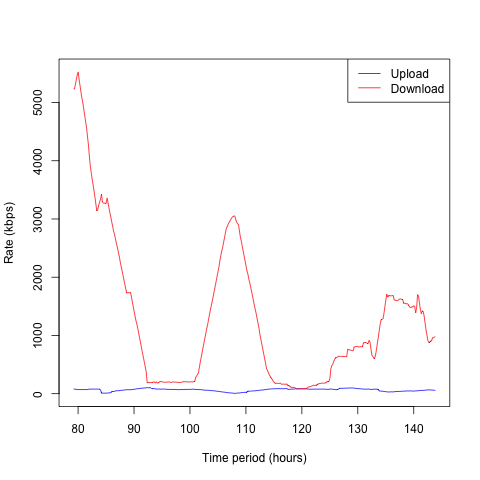}
	\end{subfigure}

	\begin{subfigure}[b]{0.24\textwidth}
		\includegraphics[width=\textwidth]{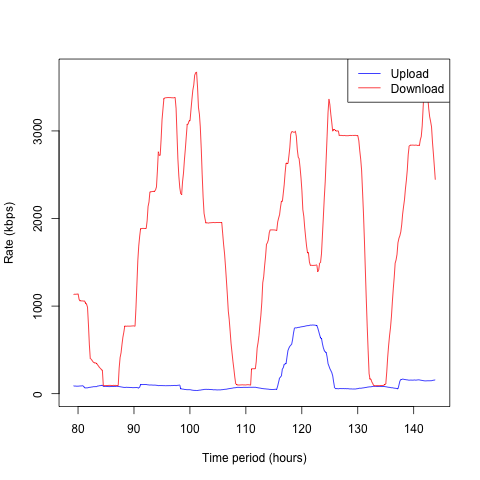}
	\end{subfigure}
	\begin{subfigure}[b]{0.24\textwidth}
		\includegraphics[width=\textwidth]{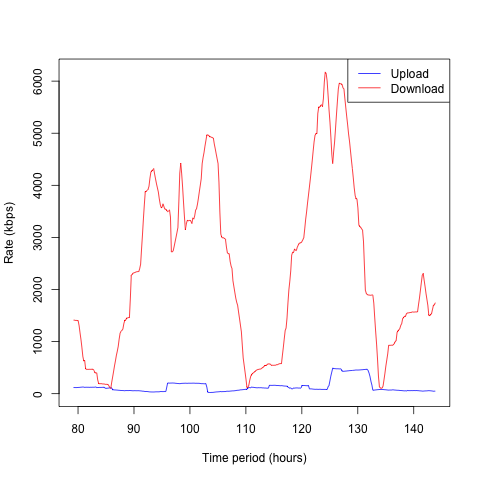}
	\end{subfigure}
	\begin{subfigure}[b]{0.24\textwidth}
		\includegraphics[width=\textwidth]{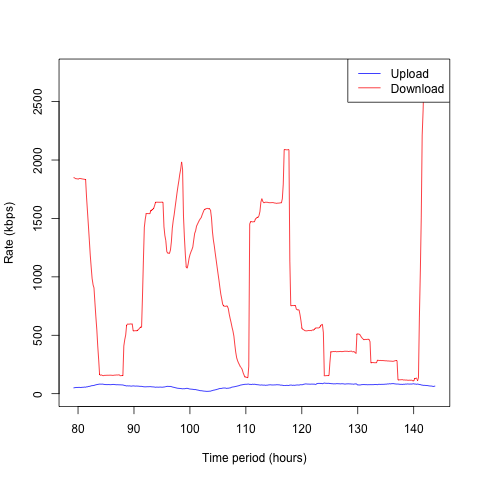}
	\end{subfigure}
	\begin{subfigure}[b]{0.24\textwidth}
		\includegraphics[width=\textwidth]{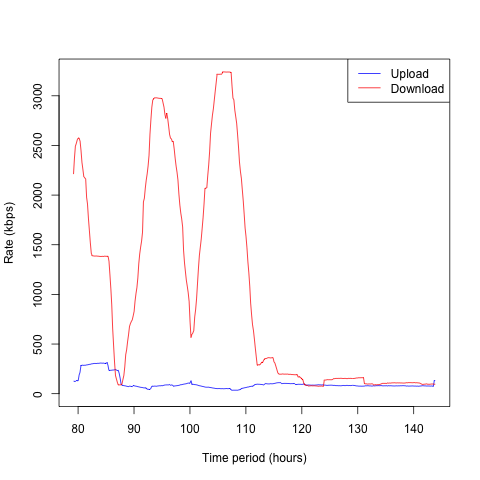}
	\end{subfigure}
	\caption{Workloads}
	\label{workloadstreams}
\end{figure*}

The individual traces can be seen in Figure~\ref{workloadstreams}.

We note that the workloads are heavily dominated by downloads, but there
are periods of high upload traffic in some workloads. To simplify
workload replay in our testbed (next section)
we only include the rate that is highest (download or upload) in each time slot,
and introduce a random noise workload in periods that have no measured traffic
in the order of 100kbps. These, volume-wise, very small adjustments do not 
change the general patterns of download and upload heavy traffic over time.

\begin{figure}[htbp]
        \centerline{\includegraphics[scale=0.5]{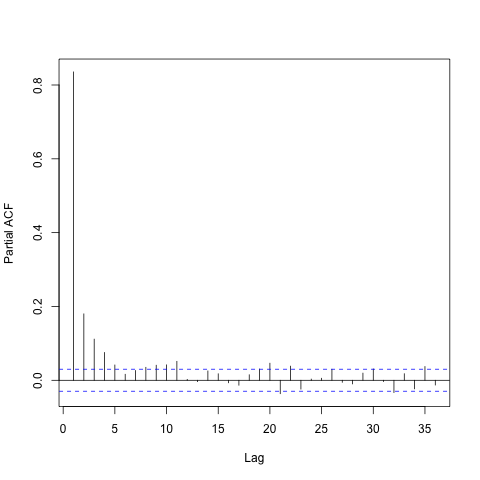}}
\caption{Partial Autocorrelation of rates with minute lags.}
\label{acf}
\end{figure}

Given that our models assume high correlation in rates between subsequent
time slots we also verified that autocorrelations are
high ($0.84$) with 1-minute lag as shown in Figure~\ref{acf}.

\section{Testbed}

To more accurately capture the effect of different
AP-STA associations on system throughput given different workloads,
we set up a testbed with 4 APs and 8 STAs powered by Raspberry Pis,
in a RF-shielded tent. 

\begin{figure}[htbp]
        \centerline{\includegraphics[scale=0.12]{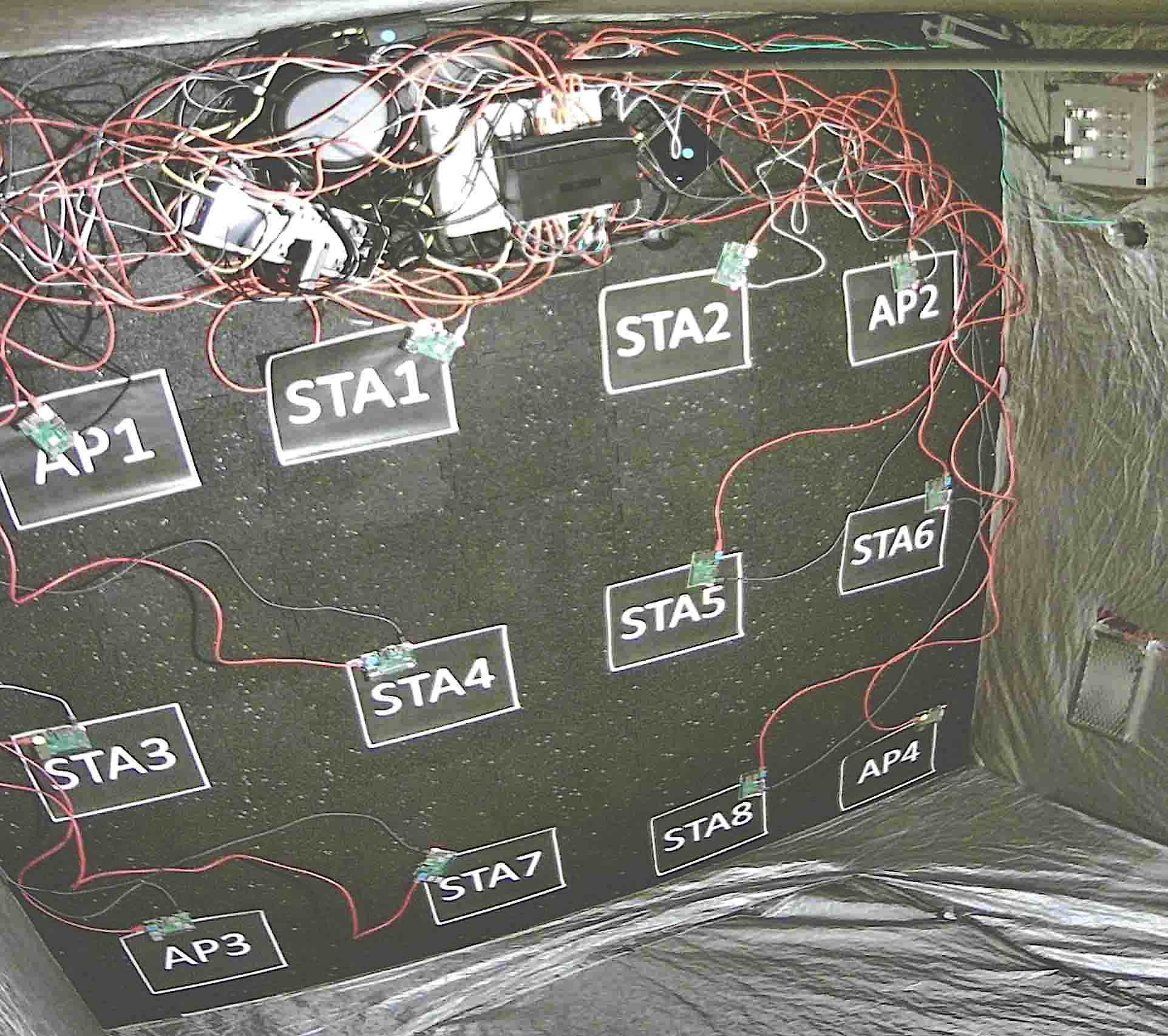}}
\caption{Testbed in RF-tent.}
\label{testbed}
\end{figure}

The tent is 7x7x7 feet large, and the APs and STAs were positioned
in a 4x3 grid with APs in all the corners (see Figure~\ref{testbed}).

The APs were set up with hostapd and dnsmasq using the internal Raspberry Pi 3 B+
W-Fi card, operating in 802.11ac (VHT) mode with channels 149, 153, 157, and 161 on
20Mhz wide bands. Maximum throughput on a single link were about 70-80Mbps with no
significant differences between positions of the APs and STAs in the tent. In other
words, all allocations are feasible in the experiments.

Both APs and STAs were configured to transmit at minimal power (1mW) to minimize interference.

The APs limit their upload and download capacities using tc qdisc htb bandwidth shaping.
Since shaping on the receiving end is limited to dropping packets which would negatively impact
throughput, we always shape on the transmitting end. That is the download capacity is set on
the AP side and the upload capacity is set on the STA side (based on the AP that the STA is
connected to).

The control plane comprises Ethernet links within a tesbed LAN from a fiber switch. Remote control
of the testbed is done via a fiber link that is pulled through a tent sleeve to avoid breaking the
Faraday property.

A Mini PC NUC connected to the same LAN controls all the experiment runs and
the AP and STA configurations over SSH links.

Traffic is replayed using nuttcp with dedicated servers on the APs for each STA, both for the 
control and data paths. All traffic is using the UDP protocol to achieve max PHY layer throughput.

The minute-by-minute workloads are replayed in parallel across all STAs. A new allocation
is enforced for each minute, and each benchmark is run for each minute until the next
minute is replayed.

The testbed is in this context used to extract throughput data across many different rate
and type (upload, download) combinations as the ground truth. The actual algorithm training and tuning is done
offline.

\section{Backhaul Constrained Experiments}
As a first step in evaluating our approach, we constrain the AP backhaul capacities as seen
in Table~\ref{T:apcapacity}.

\begin{table}[htbp]
\caption{AP Backhaul Bandwidth Capacity}
\begin{center}
\begin{tabular}{|c|c|c|}
\hline
	 & \textbf{Upload Limit} & \textbf{Download Limit}\\
	 & \textbf{(Mbps)} & \textbf{(Mbps)}\\
\hline
	\textbf{AP1 (HULD)} & $70$ & $1$  \\
	\textbf{AP2 (LUHD)} & $1$ & $70$  \\
	\textbf{AP3 (LUHD)} & $1$ & $70$  \\
	\textbf{AP4 (HULD)} & $70$ & $1$  \\
\hline
\end{tabular}
\label{T:apcapacity}
\end{center}
\end{table}

Two APs are given High Upload and Low Download (HULD)
capacities and the other two APs are given Low Upload and High Download (LUHD)
constraints. The intuition here is that biasing the AP towards serving only one
type of heavy workload has been shown to be beneficial as we mentioned in
the introduction.

Now, to test the general setup we first look at a sample set of workloads with
the behavior shown in  Table~\ref{T:stacapacity}.

\begin{table}[htbp]
\caption{Sample STA Behavior}
\begin{center}
\begin{tabular}{|c|c|c|}
\hline
\textbf{Workload} & \textbf{Type} & \textbf{Rate (Mbps)}\\
\hline
	\textbf{1 (HU)} & Upload & $50$  \\
	\textbf{2 (HD)} & Download & $50$  \\
	\textbf{3 (HD)} & Download & $50$  \\
	\textbf{4 (LU)} & Upload & $0.3$  \\
	\textbf{5 (LU)} & Upload & $0.3$  \\
	\textbf{6 (HU)} & Upload & $50$  \\
	\textbf{7 (LD)} & Download & $0.3$  \\
	\textbf{8 (LD)} & Download & $0.3$  \\
\hline
\end{tabular}
\label{T:stacapacity}
\end{center}
\end{table}

Note, we have two STAs each of the classes High Upload (HU), High Download (HD), Low Upload (LU), and Low Download (LD). 

We now define three allocations, one that mixes high upload and high download rates on the same AP (HUHD),
one that puts high rates and low rates on the same AP and keeps types together (HULU),
and finally one that mixes high rates with high rates and keeps types together (HUHU)~\footnote{one AP has HUHU the others HDHD,LDLD, and LULU}.

Table~\ref{T:sampleworkload} shows the results for these allocations.

\begin{table}[htbp]
\caption{Sample Workload Results}
\begin{center}
\begin{tabular}{|c|c|c|}
\hline
\textbf{Allocation} & \textbf{Throughput} & \textbf{Improvement}\\
 & \textbf{(Mbps)} & \textbf{over HUHD (\%)}\\
\hline
	\textbf{HUHD} & 103 & $0$  \\
	\textbf{HULU} & 197 & $91$  \\
	\textbf{HUHU} & 123 & $19$  \\
\hline
\end{tabular}
\label{T:sampleworkload}
\end{center}
\end{table}

As we can see, mapping the right set of workloads to the right AP can have
great benefits (91\% throughput improvement in this example). Moreover,
this example also shows that there could be benefits to collecting
the same type of workload on the same AP (20\% throughput improvement). 

\subsection{Trace Prediction}
We now move on to experiments with our real workload trace. 
We evaluate the approach by
first fitting the models during a training phase and then executing the
models repeatedly during a verification phase.

To be able to cover a longer
time period, which is needed to train our model, and to complete the experiment
in a reasonable time, we limit the experiment to three alternative allocations. 
The feasible set of allocations we pick are the same as the ones used with the
sample workload. Now, the rates are not static anymore so
we rename the allocations to avoid confusion. The HUHD
allocation is also the one that allocates the closest AP to each
STA so we call it the SINR allocation. The other two we call
BENCH1 and BENCH2. Note, the actual allocation is not of interest
here, simply that they are different enough to be able to expose
different system throughput behavior. 

We recall that the workload comprises 72 hours of upload and download rates,
so across the 8 workloads a total of 34560 rates are replayed across the three
allocation alternatives.

Now after collecting throughput values across all these rates and allocations
we evaluate our models as follows. 

We train a model to recognize hidden demand and predict the
optimal allocation assuming we only have observations from
a single allocation. In this case we have a clear split between
a test phase where the models are built and an
evaluation phase where we execute the model on live data.
We hence also study making this split differently, in other
words we look at how much training data we need to achieve
a certain level of performance in the predictions.

\begin{figure}[htbp]
        \centerline{\includegraphics[scale=0.5]{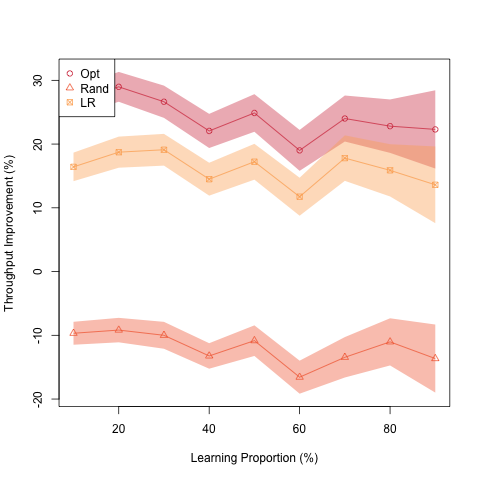}}
	\caption{Improvement over SINR allocation for backhaul constrained system.}
\label{learn_window_training}
\end{figure}

\begin{table}[htbp]
\caption{Percent Improvement over SINR Association}
\begin{center}
\begin{tabular}{|c|c|c|c|}
\hline
	\diagbox{\textbf{Model}}{\textbf{Test}} & \textbf{  10\%  } & \textbf{  30\%  } & \textbf{  90\%  }\\
\hline
	\textbf{Optimal} & $27$ & $27$  & $22$\\
	\textbf{Random} & $-10$ & $-10$ & $-14$ \\
	\textbf{LR} & $16$ & $19$ & $14$ \\
\hline
\end{tabular}
\label{T:learn_training}
\end{center}
\end{table}

The results are depicted in Figure~\ref{learn_window_training} and summarized in
Table~\ref{T:learn_training}.
With 30\% training data we observe 72\% of optimal throughput with our method.

\section{Airtime Constrained Experiments}
The next set of experiments test the proposed algorithm in the case
of hetereogeneous airtime bandwidth capacity on different Wi-Fi channels.

We constrain the AP capacities as seen
in Table~\ref{T:aircapacity} and set the workload
demands according to Table~\ref{T:staairdemand}.

\begin{table}[htbp]
\caption{AP Airtime Bandwith Capacity}
\begin{center}
\begin{tabular}{|c|c|c|}
\hline
	 & \textbf{Channel Width} & \textbf{Measured Max}\\
	 & \textbf{(Mhz)} & \textbf{(Mbps)}\\
\hline
	\textbf{AP1 (L)} & $20$ & $48$  \\
	\textbf{AP2 (H)} & $80$ & $144$  \\
\hline
\end{tabular}
\label{T:aircapacity}
\end{center}
\end{table}

\begin{table}[htbp]
\caption{Sample STA Airtime Bandwidth Demand}
\begin{center}
\begin{tabular}{|c|c|}
\hline
\textbf{Workload} & \textbf{Rate (Mbps)}\\
\hline
	\textbf{1 (H)} & $36$  \\
	\textbf{2 (L)} & $12$  \\
	\textbf{3 (H)} & $36$  \\
	\textbf{4 (H)} & $36$  \\
	\textbf{5 (L)} & $12$  \\
	\textbf{6 (L)} & $12$  \\
	\textbf{7 (H)} & $36$  \\
	\textbf{8 (L)} & $12$  \\
\hline
\end{tabular}
\label{T:staairdemand}
\end{center}
\end{table}

The SINR allocation maps all H workloads to AP1 and all L workloads to AP2,
BENCH01 maps all L workloads to AP1 and all H workloads to AP2, and finally
BENCH02 maps half of the L workloads and half of the H workloads AP1, and the
remaining workloads to AP2.

Table~\ref{T:airsampleresult} shows the results for these allocations.

\begin{table}[htbp]
\caption{Sample Airtime Results}
\begin{center}
\begin{tabular}{|c|c|c|}
\hline
\textbf{Allocation} & \textbf{Throughput} & \textbf{Improvement}\\
 & \textbf{(Mbps)} & \textbf{over SNR (\%)}\\
\hline
	\textbf{SNR} & 95 & $0$  \\
	\textbf{BENCH01} & 170 & $80$  \\
	\textbf{BENCH02} & 140 & $48$  \\
\hline
\end{tabular}
\label{T:airsampleresult}
\end{center}
\end{table}

We can see that we can improve the SNR allocation throughput with
about $80$\% with perfectly matched workloads to capacities.

\subsection{Trace Prediction}

We now replay our 3-day trace and assume that we only have access
to the throughput values observed with an SINR allocation in the
previous time slot to make an allocation decision for the next time slot.
Recall that each time slot is one minute. To saturate the airtime capacity
we aggregate data from 21 modems into 8 streams, only use two APs, and
artificially increase the traffic volume by a factor of 5. This ensures
that airtime is under contention while not changing the dynamics of
the trace. Furthermore, we only consider download replay, as uploads
are insignificant compared to the airtime used by downloads.

\begin{figure}[htbp]
        \centerline{\includegraphics[scale=0.5]{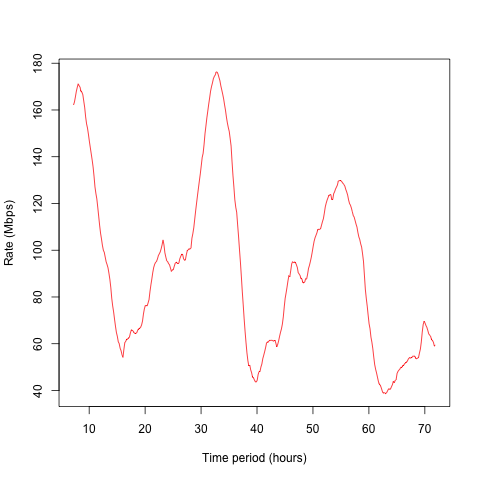}}
	\caption{Aggregate Average Download Rates over 10min Periods for Airtime Experiments.}
\label{download_band}
\end{figure}

The resulting workload can be seen in Figure~\ref{download_band}.

\begin{figure}[htbp]
        \centerline{\includegraphics[scale=0.5]{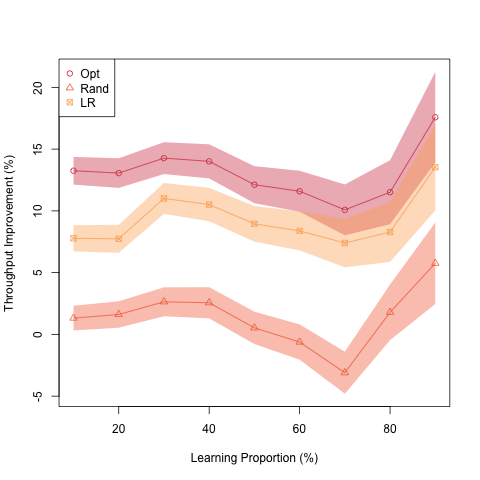}}
	\caption{Improvement over SNR allocation for airtime constrained system.}
\label{learn_band}
\end{figure}

The results with training data portions of 10-90\% are shown in Figure~\ref{learn_band} and 
summarized in Table~\ref{T:learn_band_summary}.

\begin{table}[htbp]
\caption{Percent Improvement over SNR Association}
\begin{center}
\begin{tabular}{|c|c|c|c|}
\hline
	\diagbox{\textbf{Model}}{\textbf{Test}} & \textbf{10\%} & \textbf{30\%} & \textbf{90\%}\\
\hline
	\textbf{Optimal} & $13$ & $14$ & $18$  \\
	\textbf{Random} & $3$ & $3$ & $9$  \\
	\textbf{LR} & $8$ & $11$ & $14$  \\
\hline
\end{tabular}
\label{T:learn_band_summary}
\end{center}
\end{table}

It takes more data to make the model perform and the benefit is slightly
less compared to the optimal improvement compared to the backhaul
experiment results.
With 30\% training data we observe 77\% of optimal throughput with our method.

\section{Opportunity Simulation}
We have shown that we can train our predictive models to
perform close to the optimal allocation. But what is the
potential improvement in throughput of an optimal allocation?

To find out we compare our experiment results to a simulation
replicating the network configuration of the airtime experiment.
The simulator was implemented in ns3 and use the exact same workload
data as the experiment as input.
As in the testbed, two orthogonal channels, 20Mhz and 80Mhz, were configured.
The 20Mhz channel was configured with standard {\it 80211g}
and the {\it MinstrelWifiManager} and the 80Mhz channel with the 
{\it 80211ac} standard, {\it ConstantRateWifiManager}
and DataMode and ControlMode {\it HtMcs7}. 
The 8 STAs are, like in the experiments, mapped uniformly to the two APs
and the 4 STAs assigned to an AP are positioned equidistance in a 2m radius cicrcle
around the AP with the {\it ConstantPositionMobilityModel}. The distance
was set to allow each individual STA to transfer at the max PHY rate to
the AP (if none of the other STAs transmit at the time).

We now evaluate the potential improvement in throughput by picking the
most optimal allocation in each time period, a random allocation in each time period,
and the static allocation across all periods that performs best in aggregate.

We compare the experiment result with the simulation result, and we also add more
allocations to the simulations to get closer to the true optimal allocation. Recall
that the experiment only picked 3 different allocations. In theory there are
${8 \choose 4}=70$ possible balanced allocation permutations. To restrict the
permutations further we also filter out the allocations that are reflections
of each other (same groups of STAs allocated to the APs but just on different APs).
The resulting number of permutations are then $35$. So the simulation experiment
that measures the opportunity for an optimal allocation tests 35 allocations for
each time step.

The experiment results are denoted with {\it EXP}, the simulation that uses 3 allocations
like the experiment with {\it SIM} and the simulation using all 35 allocations with {\it SIM35}.
The metrics used are {\it BEST}, denoting the improvement over the best performing static allocation,
and {\it RAND}, denoting the improvement over a method picking a random allocation in each time period.
The latter is implemented as a round robin allocator to yield deterministic simulation results, but the 
semantics is the same. Both of these measures are computed as $(V-M)/M$ where V is the studied throughput
value and M is the corresponding value for the measure (BEST or RAND).   

The throughput improvements are summarized in Table\ref{T:opportunity_summary}.

\begin{table}[htbp]
\caption{Througput Improvement Summary}
\begin{center}
\begin{tabular}{|c|c|c|}
\hline
 & \textbf{BEST} & \textbf{RAND} \\ 
\hline
	\textbf{EXP} & $0.10$ & $0.12$ \\
	\textbf{SIM} & $0.20$ & $0.26$  \\
	\textbf{SIM35} & $0.23$ & $0.37$  \\
\hline
\end{tabular}
\label{T:opportunity_summary}
\end{center}
\end{table}

The absolute throughput values between the experiment and the two simulations cannot be compared
directly as the settings are not identical, but we can compare the simulations that only differ
in the number of allocations used. Although the improvement over the best allocator is slightly
bigger in the case with only 3 possible allocations, the improvement in absolute throughput values
when using 35 allocations is about $8$\%. We also note that the improvement over the best
static allocation is a theoretical value as you would not know what the best allocation is before you
know the future workload rates. The number is just used to indicate that there is an opportunity to
dynamically change the allocation over time even for already associated streams.
The random allocation improvement is a better indicator of the true improvement opportunity. Similarly
choosing a random allocation out of a large number of potential allocations is a better indicator
of the true opportunity. Hence, the conclusion from these simulations is that we can see up to 37\%
improvement in average per-minute throughput with a density of 4 STAs per AP when load balancing across two APs using
a realistic 3-day-long cable modem trace by simply moving individual STAs between APs based on their traffic.   

\section{Calibration Simulation}
The preceding experiments showed that we can train
a model to accurately predict the best STA-AP associations
based on observed rates. Given that we need to train or calibrate
a large number of models, and that the performance during calibration will
not be better than a random allocation,
it is instrumental that the calibration is as efficient as possible,
i.e. the calibration periods are minimized while allowing the model
to be reused as long as possible.

To quantify the model performance under calibration we run simulations,
where we train 9 models (each of the three benchmark allocations both
as observed allocation and allocation to predict). The calibration
cycles through a set of calibration allocations, as follows:
\begin{equation*}
 CC = [1,1,2,1,3,2,2,3,3,1]
\end{equation*}
where $CC$ is a single calibration cycle of allocations.
When running the system through these 10 states we are able to train
each of the 9 models with a single feature list and prediction pair.

Now the question is how many cycles do we need to train with before
getting a significant improvement in throughput and at what point
does the throughput deteriorate as the calibration time performance
degrades the overall benefit of the predictions.

\begin{figure}[htbp]
        \centerline{\includegraphics[scale=0.5]{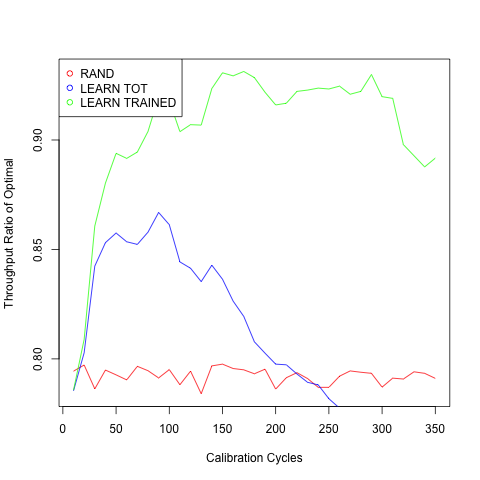}}
	\caption{Throughput as a ratio of optimal throughput given different calibration cycle lengths.}
\label{calibration}
\end{figure}

The results can be seen in Figure~\ref{calibration}.

We note that the system reaches its optimal around 100 calibration cycles. The model trained can then be used
for the remainder of the data set without training (total dataset has about 420 cycles).
The {\it LEARN TRAINED} curve shows the model performance after the model was trained. Here we see an optimal point
around 160 cycles. This indicates that there is a potential to get further improvements if the calibration
is shortened or the reuse of the model increased.

Table~\ref{T:calibrationsummary} summarizes the results.

\begin{table}[htbp]
\caption{Calibration Performance}
\begin{center}
\begin{tabular}{|c|c|c|c|}
\hline
	\diagbox{\textbf{Model}}{\textbf{Cycles}} & \textbf{10} & \textbf{100} & \textbf{170}\\
\hline
	\textbf{Random} & $80$ & $80$ & $80$  \\
	\textbf{Overall Learned} & $79$ & $86$ & $80$  \\
	\textbf{Learned after Training} & $79$ & $92$ & $93$  \\
\hline
\end{tabular}
\label{T:calibrationsummary}
\end{center}
\end{table}

In summary, the simulation shows that the calibration overhead
only reduces the improvement by 6-7 percent points compared to 
the maximum improvement during the trained regime.

\section{Implementation}
We implement the proposed method using the hostapd socket control protocol~\footnote{https://w1.fi/wpa\_supplicant/devel/hostapd\_ctrl\_iface\_page.html}.
For two APs the {\it DISASSOCIATE} command is used to move STAs between
APs. For more than two APs {\it BSS\_TM\_REQ}\footnote{BSS Transition Management Requests as defined in 802.11v} 
could be used to allow
candidate lists to steer the STA to the correct AP.
Upload and download rates are collected using {\it STA-FIRST} and {\it STA-NEXT} iterations.
These rates are used both to train the STA workload predictors
as well as to measure the aggregate system throughput. 
Python REST (Flask) servers deployed on the APs use the local hostapd socket protocol
and exposes a JSON interface to the central controller. The central controller
is also implemented in Python and uses the scikit-learn linear models package to create models,
fit models to data, and finally to predict optimal allocations (throughputs) with the fitted models.

The controller implements the following steps:
\begin{enumerate}
\item {{\bf Generate candidate set of allocations.}
       Given the number of APs and STAs available in the system
       we can enumerate all possible allocation permutations ($a^s$). Where
       $a$ is the number of APs and $s$ the number of STAs.
       To limit the potential candidate set we could filter out allocations
       that are not balanced, or in the case of two APs are reflections of each other.
       We could filter the candidate set further by only considering STAs with
       a traffic volume or SNA above a certain threshold. As an example, if we
       have two APs and we only want balanced allocations we will have $s \choose s/2$
       possible allocations. Now, given all these permutations
       we randomly select a subset. The size of this subset is a configuration parameter.
       The more samples the more likely the predicted allocation is to improve throughput,
       but the longer it takes to calibrate.
\item {{\bf Create calibration cycle of allocation transitions.} Next, we create a calibration
       cycle comprised of an ordered list of candidate allocations from the previous set. The goal
       is to train $A^2$ models, where $A$ is the number of candidate allocations. The calibration
       cycle starts with training the model ${0,0}$, i.e. the observed rates are taken from the first 
       candidate allocation and the throughput is measured in the next time step with the first allocation.  
       All possible transitions are then enumerated while minimizing the overall size of the calibration cycle.
       This would typically result in a list that has $A^2+1$ elements.}
\item {{\bf Train models.} Next, we simply loop through the
       calibration cycle a given number of iterations, adding one new feature and response pair to each model
       for each iteration. The number of cycles can be seen as the memory of the models, and again the more cycles
       the longer the calibration takes, but the more accurate the predictions might be.}
\item {{\bf Predict next allocation.} After the calibration is done we are ready to start predicting the optimal
      allocation in the next time step using our models. The currently enforced allocation as well as the STA
      rates are collected and fed into all the models that have the allocation as the starting state. The model
      corresponding to the allocation with the highest predicted throughput is then selected and enforced. The newly 
      enforced allocation then becomes the starting state for the next prediction.}
\item {\bf Update models.} Every time we make a prediction we also collect the ground truth of the obtained throughput
      given an observed state with one allocation and an enforced allocation. The model corresponding to this state transition
      is then updated by simply popping the oldest feature and response pair and adding the new one, to maintain the length
      of the memory from the calibration. Alternatively and arbitrary memory size may be specified where popping does not
      start until a threshold is reached. The purpose of that would be to allow predictions before the system is fully calibrated,
		but the downside is that there may then be an uneven number of training data points (memory) in the different models during an interim period.}
\end{enumerate}
The first three steps are typically only done during a bootstrap phase or
after there has been a significant change in the stations associated with the
APs, and we hence refer to them as the calibration phase.

Now, what happens if a new STA enters the system or an old one drops out? In the first case we could simply ignore
the new STA in our models and just have the newly measured throughput be an indirect indicator of a new STA impacting
the optimal allocation. However, in that case we cannot move the new STA which may or may not be an issue.
Alternatively, we could add the new STA to all the existing trained models and simply add a 0 for the historical rates
of the new STA. If a STA drops out the rates would just naturally all become 0 and the impact of the STA in the model
will gradually diminish. The STA would however still have a model parameter that consumes computer memory and, hence
it could similarly to the new STA at some point also be explicitly deleted from the model. To make this work, we not only keep
the fitted model in memory but also the feature and response arrays so we can easily refit modified feature arrays.

\section{Analysis}
To improve our intuition why the proposed method works, we illustrate with the
simplest possible case of two APs and two STAs. Let's also assume that
the first AP has higher bandwidth capacity than the first AP, e.g. uses
a wider band or a higher frequency.

The total number of permutations is $2^2=4$ and the number of balanced allocations are 
${2 \choose 1} = 2$. The allocations are $A_1={1,2}$ and $A_2={2,1}$, i.e. put $s_1$ on $a_1$ and
$s_2$ on $a_2$ or $s_1$ on $a_2$ and $s_2$ on $a_1$. 
The calibration cycle then becomes $[1,1,2,2,1]$ to capture all transitions and train all four linear models,
which are (we remove intercept and error terms for clarity):
\begin{equation}
\begin{aligned}
m_{A_1,A_1} = k_1 r_1 + k_2 r_2 \\
m_{A_1,A_2} = k_3 r_1 + k_4 r_2 \\
m_{A_2,A_1} = k_5 r_1 + k_6 r_2 \\
m_{A_2,A_2} = k_7 r_1 + k_8 r_2
\end{aligned}
\end{equation}
where $m_{A_1,A_2}$ is the model predicting the throughput of allocation $A_2$ given rates $r$
observed with allocation $A_1$, and $k$ represents the model coefficients to be fit.

Now, given that $a_1$ has more bandwidth than $a_2$, and that we are currently in a state $A_1$, 
we can expect that the throughput for $s_1$ would stay the same or drop in a transition to $A_2$,
and vice versa, the throughput for $s_2$ would increase or stay the same in a transition to $A_2$. 

So if the average observed $r_1$ is higher in state $A_1$ than in $A_2$ then $k_1 >k_3$.
Similarly if the average observed $r_2$ is lower in state $A_1$ than in $A_2$ then $k_2 < k_4$.
Hence, $m_{A_1,A_1}$ should yield a higher throughput than $m_{A_1,A_2}$ for an
observed rate of $r_1$ and $r_2$, and in other words the optimal next allocation
is to put $s_1$ on $a_1$ and $s_2$ on $a_2$, which corresponds to intuition.

\section{Conclusions}

In summary, we have verified experimentally with real workload traces that 
different associations of STAs with APs based on transmission rates
can improve throughput
significantly compared to SINR associations in a dense Wi-Fi
network. 

A simple linear regression ensemble model
shows good performance when learning hidden demand
and predicting optimal allocations in future
time periods.

\section*{Acknowledgements}
We thank Lin Cheng, Hang Ung and John Bahr for helpful feedback on
early drafts of this paper.

\bibliographystyle{ACM-Reference-Format}
\bibliography{related}
\appendix
\end{document}